\documentclass[12pt]{article}
\usepackage{amsfonts}

\usepackage{amsmath}
\usepackage{graphicx}

\input{tcilatex}
\begin{document}

\title{Physical coordinates as dynamic variables for the superparticle from
its geometrical action}
\author{Diego Julio Cirilo-Lombardo \\
%EndAName
{\small Bogoliubov Laboratory of Theoretical Physics}\\
{\small Joint Institute of Nuclear Research, 141980, Dubna, Russian
Federation.}\\
{\small e-mails: diego@thsun1.jinr.ru ; diego77jcl@yahoo.com}}
\maketitle

\begin{abstract}
We show that the particle actions in the superspace that are invariant with
respect to general covariance transformations can be formulated in terms of
physical coordinates with non zero evolution Hamiltonians by identifying
these coordinates with some dynamic variables. The local $\kappa $-symmetry
for superparticle actions in this formulation is briefly discussed.


\end{abstract}

\section{Introduction and summary}

\bigskip

It is known [1]-[5] that the reparameterization invariance of the theories
of relativistic particles and relativistic strings, as well as the
invariance of the gravity theory with respect to general covariance
transformations, results in serious problems when analyzing these theories
in the Hamiltonian formalism : occurs the nullification of the Hamiltonian
because of this invariance.

For the relativistic particle, this problem was already circumvented in the
first Einstein and Poincare' papers [6,7] thanks to identifying one of the
dynamical variables, $x_{0}\left( \tau \right) $ , with the physical time,
and this identification is natural in the special relativity theory.

The idea that the time measured by an observer's clock is a dynamic variable
seems rather strange in field theories, where the fields are functions of
space and time. In the string theory, time was considered a dynamic variable
in [8, 9], which allows relating this theory with the Born-Infeld theory.

It is possible to give a correct description of these reparameterizations
invariant systems, from a dynamical point of view, passing to the physical
variables by means of the Levi-Civita canonical transformations, as was
shown in [10, 11]. These canonical transformations make the dynamical system
under consideration suitable to be integrated or quantized.

Strongly motivated to extend the above concepts to a toy superspace we apply
the Levi-Civita canonical transformations to the simple model of
superparticle of Volkov and Pashnev [12,13], that is the type G4 in the
description of Casalbuoni [14,15] to obtain the unconstrained form of the
superparticle action, after that these canonical transformations have been
performed. This final unconstrained action, that is in function on the
physical variables, is suitable to be quantized or integrated. Recently in
[18, 19], the importance of the constraints in the superparticle actions
when local supersymmetry transformations are introduced, was shown. The
space-time covariant formulation of super p-branes is known to have a local
fermionic invariance on the world manifold, first discovered by Siegel [21]
for the superparticle and posteriorly in [22] for superstrings. This
invariance helps to balance the number of commuting and anti-commuting
degrees of freedom mainly in models where with the boson and fermion
variables belonging to different representations of the Lorentz group of the
target space-time. As the parameter of this transformation is an
anti-commuting space-time spinor $\kappa $ varying in an arbitrary way over
the world manifold. In this sense this $\kappa $-invariance is a
supersymmetry. The another motivation of this paper is to discuss shortly
what happens with the global SUSY\ and this local $\kappa $-supersymmetry
when these Levi-Civita canonical transformations are performed in the
superparticle-model under consideration because it is well established that
the actions that are $\kappa $-invariants have the physical interpretation
as the leading term in the effective action describing the the low energy
dynamics of topological defects of supersymmetric field theories [23-28].

\section{The superparticle model}

\noindent In the superspace the coordinates are given not only by the
spacetime $x_{\mu }$ coordinates, but also for anticommuting spinors $\theta
^{\alpha }$ and $\overline{\theta }^{\overset{.}{\alpha }}$ . The resulting
metric [12,13] must be invariant to the action of the Poincare group, and
invariant also to the supersymmetry transformations
\begin{equation*}
x_{\mu }^{\prime }=x_{\mu }+i\left( \theta ^{\alpha }\left( \sigma \right)
_{\alpha \overset{.}{\beta }}\overline{\xi }^{\overset{.}{\beta }}-\xi
^{\alpha }\left( \sigma \right) _{\alpha \overset{.}{\beta }}\overline{%
\theta }^{\overset{.}{\beta }}\right) \ ;\ \ \theta ^{\prime \alpha }=\theta
^{\alpha }+\xi ^{\alpha }\ ;\ \ \overline{\theta ^{\prime }}^{\overset{.}{%
\alpha }}=\overline{\theta }^{\overset{.}{\alpha }}+\overline{\xi }^{\overset%
{.}{\alpha }}
\end{equation*}

The simplest super-interval that obeys the requirements of invariance given
above, is the following
\begin{equation}
ds^{2}=\omega _{\mu }\omega ^{\mu }+a\omega _{\alpha }\omega ^{\alpha
}-a^{\ast }\omega _{\overset{.}{\alpha }}\omega ^{\overset{.}{\alpha }}
\end{equation}%
where
\begin{equation*}
\omega _{\mu }=dx_{\mu }-i\left( d\theta \ \sigma _{\mu }\overline{\theta }%
-\theta \ \sigma _{\mu }d\overline{\theta }\right) ;\ \ \ \ \ \ \ \ \omega
^{\alpha }=d\theta ^{\alpha }\ ;\ \ \ \ \ \ \ \ \omega ^{\overset{.}{\alpha }%
}=d\overline{\theta }^{\overset{.}{\alpha }}
\end{equation*}%
are the Cartan forms of the group of supersymmetry [17].

The spinorial indexes are related as follows
\begin{equation*}
\theta ^{\alpha }=\varepsilon ^{\alpha \beta }\theta _{\beta }\ \ \ \ \ ;\ \
\ \ \ \theta _{\alpha }=\theta ^{\beta }\varepsilon _{\beta \alpha }\ \ \ ;\
\ \varepsilon _{\alpha \beta }=-\varepsilon _{\beta \alpha }\ \ \ \ ;\ \ \
\varepsilon ^{\alpha \beta }=-\varepsilon ^{\beta \alpha }\ \ \ ;\ \ \
\varepsilon _{12}=\varepsilon ^{12}=1
\end{equation*}
and of analog manner for the spinors with punctuated indexes. The complex
constants $a$ and $a^{\ast }$ in the line element (1) are arbitrary. This
arbitrarity for the choice of $a$ and $a^{\ast }$are constrained by the
invariance and reality of the interval (1).

As we have extended our manifold to include fermionic coordinates, it is
natural extend also the concept of trayectory of point particle to the
superspace. To do this we take the coordinates $x\left( \tau \right) $, $%
\theta \left( \tau \right) $ and $\overline{\theta }^{\overset{.}{\alpha }%
}\left( \tau \right) $ depending on the evolution parameter $\tau .$
Geometrically, the function action that will describe the world-line of the
superparticle, is
\begin{equation}
S=-m\int_{\tau 1}^{\tau 2}d\tau \sqrt{\overset{\circ }{\omega _{\mu }}%
\overset{\circ }{\omega ^{\mu }}+a\overset{.}{\theta _{\alpha }}\overset{.}{%
\theta ^{\alpha }}-a^{\ast }\overset{.}{\overline{\theta }_{\overset{.}{%
\alpha }}}\overset{.}{\overline{\theta }^{\overset{.}{\alpha }}}}=\int_{\tau
1}^{\tau 2}d\tau L\left( x,\theta ,\overline{\theta }\right)
\end{equation}
where $\overset{\circ }{\omega _{\mu }}=\overset{.}{x}_{\mu }-i\left(
\overset{.}{\theta }\ \sigma _{\mu }\overline{\theta }-\theta \ \sigma _{\mu
}\overset{.}{\overline{\theta }}\right) $ and the upper point means
derivative with respect to the parameter $\tau $, as is usual.

The momenta, canonically conjugated to the coordinates of the superparticle,
are
\begin{equation*}
\mathcal{P}_{\mu }=\partial L/\partial x^{\mu }=\left( m^{2}/L\right)
\overset{\circ }{\omega _{\mu }}
\end{equation*}
\begin{equation*}
\mathcal{P}_{\alpha }=\partial L/\overset{.}{\partial \theta ^{\alpha }}=i%
\mathcal{P}_{\mu }\left( \sigma ^{\mu }\right) _{\alpha \overset{.}{\beta }}%
\overline{\theta }^{\overset{.}{\beta }}+\left( m^{2}a/L\right) \overset{.}{%
\theta _{\alpha }}
\end{equation*}
\begin{equation}
\mathcal{P}_{\overset{.}{\alpha }}=\partial L/\overset{.}{\partial \overline{%
\theta }^{\overset{.}{\alpha }}}=i\mathcal{P}_{\mu }\theta ^{\alpha }\left(
\sigma ^{\mu }\right) _{\alpha \overset{.}{\alpha }}-\left( m^{2}a/L\right)
\overset{.}{\overline{\theta }_{\overset{.}{\alpha }}}
\end{equation}
It is difficult to study this system in the Hamiltonian formalism framework
because of the constraints and the nullification of the Hamiltonian. As the
action (2)\ is invariant under reparametrizations of the evolution parameter
\begin{equation}
\tau \rightarrow \widetilde{\tau }=f\left( \tau \right)
\end{equation}
one way to overcome this difficulty is to make the dynamic variable $x_{0}$
the time. For this, it is sufficient to use the chain rule of derivatives
(with special care of the anticommuting variables)\footnote{%
We take the Berezin convention for the Grassmannian derivatives: $\ \delta
F(\theta )=\frac{\partial F}{\partial \theta }\delta \theta $} and to write
the action in the form
\begin{equation}
S=-m\int_{\tau 1}^{\tau 2}\overset{.}{x}_{0}d\tau \sqrt{\left[ 1-iW_{,0}^{0}%
\right] ^{2}-\left[ x^{i}-W_{,0}^{i}\right] ^{2}+a\overset{.}{\theta
_{\alpha }}\overset{.}{\theta ^{\alpha }}-a^{*}\overset{.}{\overline{\theta }%
_{\overset{.}{\alpha }}}\overset{.}{\overline{\theta }^{\overset{.}{\alpha }}%
}}
\end{equation}
where the $W_{,0}^{\mu }$ was defined by
\begin{equation*}
\overset{\circ }{\omega }^{0}=\overset{.}{x}^{0}\left[ 1-iW_{,0}^{0}\right]
\end{equation*}
\begin{equation}
\overset{\circ }{\omega }^{i}=\overset{.}{x}^{0}\left[ x_{,0}^{i}-iW_{,0}^{i}%
\right]
\end{equation}
whence $x_{0}\left( \tau \right) $ turns out to be the evolution parameter
\begin{equation}
S=-m\int_{x_{0}\left( \tau _{1}\right) }^{x_{0}\left( \tau _{2}\right)
}dx_{0}\sqrt{\left[ 1-iW_{,0}^{0}\right] ^{2}-\left[ x^{i}-W_{,0}^{i}\right]
^{2}+a\overset{.}{\theta _{\alpha }}\overset{.}{\theta ^{\alpha }}-a^{*}%
\overset{.}{\overline{\theta }_{\overset{.}{\alpha }}}\overset{.}{\overline{%
\theta }^{\overset{.}{\alpha }}}}\equiv \int dx_{0}L
\end{equation}
Physically this parameter (we call it the dynamical parameter) is the time
measured by an observer's clock in the rest frame.

Therefore, the invariance of a theory with respect to the invariance of the
coordinate evolution parameter means that one of the dynamic variables of
the theory ($x_{0}\left( \tau \right) $ in this case) becomes the observed
time with the corresponding non-zero Hamiltonian
\begin{equation}
H=\mathcal{P}_{\mu }\overset{.}{x}^{\mu }+\Pi _{\alpha }\overset{.}{\theta }%
^{\alpha }+\Pi _{\overset{.}{\alpha }}\overset{.}{\theta }^{\overset{.}{%
\alpha }}-L  \notag
\end{equation}
\begin{equation}
\ \ \ \ \ \ \ \ \ \ \ \ \ \ \ \ \ =\sqrt{m^{2}-\left( \mathcal{P}_{\mu }%
\mathcal{P}^{\mu }+\frac{1}{a}\Pi _{\alpha }\Pi ^{\alpha }-\frac{1}{a^{*}}%
\Pi _{\overset{.}{\alpha }}\Pi ^{\overset{.}{\alpha }}\right) }
\end{equation}
where
\begin{equation*}
\Pi _{\alpha }=\mathcal{P}_{\alpha }+i\ \mathcal{P}_{\mu }\left( \sigma
^{\mu }\right) _{\alpha \overset{.}{\beta }}\overline{\theta }^{\overset{.}{%
\beta }}
\end{equation*}
\begin{equation*}
\Pi _{\overset{.}{\alpha }}=\mathcal{P}_{\overset{.}{\alpha }}-i\mathcal{P}%
_{\mu }\theta ^{\alpha }\left( \sigma ^{\mu }\right) _{\alpha \overset{.}{%
\alpha }}
\end{equation*}
Choosing $x_{0}\left( \tau \right) $ as the evolution parameter, we thus fix
the reference frame. This procedure to fix the reference frame is called
\textit{the physical realization of the relativistic particle} [11].When a
specific physical realization is chosen, we lose all other realizations. In
particular, the co-moving frame in which the time is the proper time with
the interval
\begin{equation}
dt=\sqrt{\frac{ds^{2}}{d\tau ^{2}}}d\tau
\end{equation}
But, it is easy to show that the relativistic action (2), written in the
invariant form with the additional variable $e\left( \tau \right) $
(einbein)
\begin{equation*}
S=-\frac{m}{2}\int_{\tau 1}^{\tau 2}d\tau \left[ \frac{1}{e}\left( \overset{%
\circ }{\omega _{\mu }}\overset{\circ }{\omega ^{\mu }}+a\overset{.}{\theta
_{\alpha }}\overset{.}{\theta ^{\alpha }}-a^{*}\overset{.}{\overline{\theta }%
_{\overset{.}{\alpha }}}\overset{.}{\overline{\theta }^{\overset{.}{\alpha }}%
}\right) +e\right]
\end{equation*}
\begin{equation}
=\int d\tau \left[ \mathcal{P}_{\mu }\mathcal{P}^{\mu }+\frac{1}{a}\Pi
_{\alpha }\Pi ^{\alpha }-\frac{1}{a^{*}}\Pi _{\overset{.}{\alpha }}\Pi ^{%
\overset{.}{\alpha }}\right] -\frac{e}{2m}\left[ m^{2}-\left( \mathcal{P}%
_{\mu }\mathcal{P}^{\mu }+\frac{1}{a}\Pi _{\alpha }\Pi ^{\alpha }-\frac{1}{%
a^{*}}\Pi _{\overset{.}{\alpha }}\Pi ^{\overset{.}{\alpha }}\right) \right]
\end{equation}
also describes the relativistic particle in the co-moving frame. The
equations of motion for the action (11) are
\begin{equation*}
m^{2}-\left( \mathcal{P}_{\mu }\mathcal{P}^{\mu }+\frac{1}{a}\Pi _{\alpha
}\Pi ^{\alpha }-\frac{1}{a^{*}}\Pi _{\overset{.}{\alpha }}\Pi ^{\overset{.}{%
\alpha }}\right) =0
\end{equation*}
\begin{equation}
\frac{\partial H}{\partial \Pi }=\overset{.}{\theta }\left( \text{or}\overset%
{.}{\overline{\theta }}\right) \ \ \ \ \ \ \ \ \ \ \ \ \ \ \ \ \ \ \ \ \ \ \
\ \frac{\partial H}{\partial \mathcal{P}}=\overset{.}{x}
\end{equation}
Equations (12)\ contain two times: $x_{0}$ is the time in the rest frame and
t is the time in the co-moving frame.

The relation
\begin{equation}
x_{0}=\frac{\mathcal{P}_{0}}{m}t
\end{equation}%
between these two times describes the purely relativistic effect of changing
the time when passing to another reference frame.

As shown in [12], there exists a scenario of a dynamic transition to the
co-moving frame using the Levi-Civita canonical transformation $\left(
\mathcal{P}_{\mu },\Pi _{\alpha },\Pi _{\alpha };x_{\mu },\theta _{\alpha },%
\overline{\theta }_{\overset{.}{\alpha }}\right) \rightarrow \left( P_{\mu
},P_{\alpha },P_{\overset{.}{\alpha }};Q_{\mu },Q_{\overset{.}{\alpha }%
},Q_{\alpha }\right) $ [10, 11]
\begin{equation}
P_{0}=\frac{1}{2m}\left( \mathcal{P}_{\mu }\mathcal{P}^{\mu }+\frac{1}{a}\Pi
_{\alpha }\Pi ^{\alpha }-\frac{1}{a^{\ast }}\Pi _{\overset{.}{\alpha }}\Pi ^{%
\overset{.}{\alpha }}\right) \ \ \ \ ,\ \ \ P_{i}=\mathcal{P}_{i}\ \ \ ,\ \
\ P_{\alpha }=\Pi _{\alpha }\ \ \ \ \ \ P_{\overset{.}{\alpha }}=\Pi _{%
\overset{.}{\alpha }},\
\end{equation}%
\begin{equation*}
Q_{0}=x_{0}\frac{m}{\mathcal{P}_{0}}\ \ \ Q_{i}=x_{i}+\frac{x_{0}}{\mathcal{P%
}_{0}}\mathcal{P}_{i}\ \ \ \ \ \ \ \ \ \ Q_{\alpha }=\theta _{\alpha }+\frac{%
x_{0}}{\mathcal{P}_{0}a}\Pi _{\alpha }\ \ \ \ \ \ \ \ \ Q_{\overset{.}{%
\alpha }}=\overline{\theta }_{\overset{.}{\alpha }}-\frac{x_{0}}{\mathcal{P}%
_{0}a^{\ast }}\Pi _{\overset{.}{\alpha }}\ \ \ \
\end{equation*}%
which transforms the constraint into the new momentum $P_{0}$ and the time $%
x_{0}$ into the proper time (10). Indeed we can use the eqs. (14)\ to
express the old momenta $\mathcal{P}_{\mu },\Pi _{\alpha },\Pi _{\alpha }$
and coordinates $x_{\mu },\theta _{\alpha },\overline{\theta }_{\overset{.}{%
\alpha }}$ through the new ones as
\begin{equation*}
\mathcal{P}_{0}=\pm \sqrt{2mP_{0}-\left( P_{i}P^{i}+\frac{1}{a}P_{\alpha
}P^{\alpha }-\frac{1}{a^{\ast }}P_{\overset{.}{\alpha }}P^{\overset{.}{%
\alpha }}\right) }\ \ \ \ ;\ \ \ \ \ \ \ \ \ \ \ \mathcal{P}_{i}=P_{i}
\end{equation*}%
\begin{equation*}
x_{0}=\pm \frac{Q_{0}}{m}\sqrt{2mP_{0}-\left( P_{i}P^{i}+\frac{1}{a}%
P_{\alpha }P^{\alpha }-\frac{1}{a^{\ast }}P_{\overset{.}{\alpha }}P^{\overset%
{.}{\alpha }}\right) }
\end{equation*}%
\begin{equation}
x_{i}=Q_{i}+\frac{Q_{0}}{m}P_{i}\ \ \ \ \ \ \ \ \ \ \theta _{\alpha
}=Q_{\alpha }+\frac{Q_{0}}{ma}P_{\alpha }\ \ \ \ \ \ \ \ \ \overline{\theta }%
_{\overset{.}{\alpha }}=Q_{\overset{.}{\alpha }}-\frac{Q_{0}}{ma^{\ast }}P_{%
\overset{.}{\alpha }}
\end{equation}%
the action (7) in the new variables then becomes
\begin{equation}
S=\int_{\tau _{1}}^{\tau _{2}}\left[ P_{\mu }\overset{.}{Q^{\mu }}+P_{\alpha
}\overset{.}{Q^{\alpha }}+P_{\overset{.}{\alpha }}\overset{.}{Q}^{\overset{.}%
{\alpha }}-e\left( P_{0}-\frac{m}{2}\right) +\frac{d\left( tP_{0}\right) }{dt%
}\right]
\end{equation}%
Varying the action (16) with respect to $P_{0}$, we define the new variable $%
Q_{0}=t$%
\begin{equation}
\frac{dQ_{0}}{d\tau }=e\left( \tau \right)
\end{equation}%
to be the new proper time (10) , and varying (16) with respect to $e\left(
\tau \right) $ we obtain the constraint
\begin{equation}
P_{0}-\frac{m}{2}=0
\end{equation}%
Finally, resolving this constraint with respect to the momentum component $%
P_{0}$, we obtain the expression
\begin{equation}
S=\int_{t_{1}}^{t_{2}}\left[ P_{i}\frac{dQ^{i}}{dt}+P_{\alpha }\frac{%
dQ^{\alpha }}{dt}+P_{\overset{.}{\alpha }}\frac{dQ^{\overset{.}{\alpha }}}{dt%
}-\frac{m}{2}-\frac{d\left( tP_{0}\right) }{dt}\right]
\end{equation}%
Inverse Levi-Civita canonical transformations (14) and solutions (16) and
(17) establish the relation between two reference frames, in this case in
the superspace, with different physical realizations of the same particle.
The reparametrization invariance therefore allows describing two physical
realizations of the same particle by two constraint-free mechanics, while
these mechanics are related through purely relativistic effects. In the
dynamic transition given by the formulas (13), the global SUSY\ is preserved
but the the $\kappa $-invariance is not explicitly manifest in the
expression (18). In order to restore the local relativistic $\kappa $%
-symmetry there are two possibilities to induce it in (18). These two ways
are\footnote{%
Here is the $\mathcal{L}$ is the Lagrangian density, $\varsigma \overset{.}{%
,\varsigma }(\overline{\varsigma },\overset{.}{\overline{\varsigma }})$ are
Lagrange multipliers and $D\left( \overline{D}\right) $are the covariant
derivatives, as usual.}%
\begin{equation*}
\mathcal{L}_{A}\equiv \mathcal{L}+\varsigma D+\overline{\varsigma D}+eH
\end{equation*}%
\begin{equation*}
\mathcal{L}_{B}\equiv \mathcal{L}+\overset{.}{\varsigma }D+\overset{.}{%
\overline{\varsigma D}}+eH
\end{equation*}%
where only the second choice (i.e:$\mathcal{L}_{B}$) is the correct one%
\footnote{%
It fact is easily seen from the point of view of the first order formulation
of the superparticle action: when the metric (1) degenerates, this new
Lagrangian is closest to the Siegel model.}. It is interesting to note that
these different ways to introduce the local supersymmetry was also obtained
and analyzed in ref.[20] taking as the starting point the functional
approach of the the classical mechanics and the BRST\ formulation where was
explicitly shown that are two gauge theories which differ in the gauge
couplings: in the Lagrangian and in the physical Hilbert spaces.

\section{Acknowledgements}

I am very grateful to Professors J. A. Helayel-Neto, B. M. Barbashov, A.
Dorokhov and E. A. Ivanov for very useful discussions. Thanks also are given
to the Directorate of the JINR, in particular of the Bogoliubov Laboratory
of Theoretical Physics, for their hospitality and support.

\section{\protect\bigskip References}

\bigskip

[1] P. A. M. Dirac, Proc. Roy. Soc. \textbf{246}, 333 (1958); Phys. Rev.
\textbf{114}, 924 (1959).

[2] R. Arnowitt, S. Deser and C. W. Misner, Phys. Rev. \textbf{116}, 1322
(1959); \textbf{117}, 1595 (1960); \textbf{122}, 997 (1961).

[3] P. A. M. Dirac, \textit{Lectures on Quantum Mechanics}, Belfer Graduate
School of Science, Yeshiva University, New York (1964).

[4] L. D. Fadeev and V. N. Popov, Sov Phys. Usp. \textbf{111}, 427 (1973).

[5] B. M. Barbashov and V. V. Nesterenko, \textit{Introduction to
Relativistic String Theory}, Energoatomizdat, Moscow (1987) [in Russian];
English transl., World Scientific, Singapore (1990).

[6] H. Poincare', Comp. Ren. Acad. Sci. Paris \textbf{140}, 1504 (1904).

[7] A. Einstein, Ann. Phys. \textbf{17}, 891 (1905).

[8] B. M. Barbashov and N. A. Chernikov, ''\textit{Classical dynamics of the
relativistic string'' [in Russian]}, Preprint P2-7852, JINR, Dubna (1974).

[9] A. G. Reiman and L. D. Fadeev, Vestn. Leningr. Gos. Univ., No. 1, 138
(1975).

[10] T. Levi-Civita, Prace Mat.-Fiz. \textbf{17}, 1 (1906); T. Levi-Civita
and U. Amaldi, Lezzione di Meccania Razionale (Nicola Zanichelli, Bologna
1927), Vol. II, Part 2.

[11] S. Shanmugadhasan, J. Math. Phys. \textbf{14}, 677 (1973).

[12] B. M. Barbashov, V. N. Pervushin and D. V. Proskurin, Theor. Math.
Phys. \textbf{132} (2), 1045 (2002).

[13] D. V. Volkov and S. V. Peletminskii, JETP \textbf{37}, 170 (1959) [in
Russian].

[14] A.I.Pashnev and D. V. Volkov, Teor. Mat. Fiz., Tom. \textbf{44}, No. 3,
321 (1980) [in Russian].

[15] R. Casalbuoni, \ Nuovo. Cim., Vol. 33\textbf{A}, N. 3, 389 (1976).

[16] R. Casalbuoni, Phys. Lett. \textbf{62}B, 49 (1976).

[17] A. P. Akulov and D. V. Volkov, Piz'ma JETF \textbf{16}, 621 (1972).

[18] P. A. Grassi, G. Policastro and P. van Nieuwenhuizen, Phys. Lett.
\textbf{B} 553, 96 (2003).

[19] P. D. Jarvis et al., Phys. Lett. \textbf{B} 427, 47 (1998).

[20] E. Deotto, Europhys. Lett. 58, 195 (2002), and references therein.

[21] W. Siegel, Phys. Lett. \textbf{B} 128, 397 (1983).

[22] M. B. Green and J. H. Schwarz, Phys. Lett. \textbf{B} 136, 372 (1986).

[23] J. Hughes, J. Liu and J. Polchinski, Phys. Lett. \textbf{B} 180 (1986).

[24] P. K. Townsend, Phys. Lett.\textbf{\ B} 202, 53 (1988).

[25] A. Achucarro, J. P. Gaunlett, K. Itoh and P. K. Townsend, Nucl. Phys.
\textbf{B} 314, 129 (1989).

[26] J. P. Gaunlett, Phys. Lett \textbf{B} 228, 188 (1989).

[27] E. A. Ivanov and \ A. A. Kapustnikov, Phys. Lett. \textbf{B} 252, 212
(1990).

[28] J. A. Azcarraga, J. P. Gaunlett, J. M. Izquierdo and P. K. Townsend,
Phys. Rev. Lett. 63, 2443 (1989).

\end{document}